\documentclass[acmsmall]{acmart}

\usepackage{listings}
\usepackage{hyperref}
\usepackage{graphicx}
\usepackage{csvsimple}
\usepackage{subcaption}
\usepackage{comment}
\usepackage{xcolor}
\usepackage{color, colortbl}
\usepackage{url}
\usepackage{multirow}
\usepackage{amsmath}
\usepackage{rotating}
\usepackage[T1]{fontenc}
\usepackage{soul}
\usepackage{tikz}
\usepackage{makecell}

\definecolor{Gray}{gray}{0.8}
\acmDOI{}
\renewcommand\footnotetextcopyrightpermission[1]{} 
\begin{document}

\title{Hybrid Human-AI Curriculum Development for Personalised Informal Learning Environments}

\author{Mohammadreza Tavakoli}
\affiliation{
\institution{Leibniz Information Centre for Science and Technology (TIB)}
\country{Germany}}
\email{reza.tavakoli@tib.eu}

\author{Abdolali Faraji}
\affiliation{
\institution{Leibniz Information Centre for Science and Technology (TIB)}
\country{Germany}}
\email{abdolali.faraji@tib.eu}

\author{Mohammadreza Molavi}
\affiliation{
\institution{Amirkabir University of Technology}
\country{Iran}}
\email{mr.molavi@aut.ac.ir}

\author{Stefan T. Mol}
\affiliation{
\institution{University of Amsterdam}
\country{Netherlands}}
\email{s.t.mol@uva.nl}

\author{G\'abor Kismih\'ok}
\affiliation{%
\institution{Leibniz Information Centre for Science and Technology (TIB)}
\country{Germany}}
\email{gabor.kismihok@tib.eu}

\renewcommand{\shortauthors}{Tavakoli and Faraji, et al.}

\begin{abstract}
Informal learning procedures have been changing extremely fast over the recent decades not only due to the advent of online learning, but also due to changes in what humans need to learn to meet their various life and career goals. Consequently, online, educational platforms are expected to provide personalized, up-to-date curricula to assist learners. Therefore, in this paper, we propose an \emph{Artificial Intelligence} (\emph{AI}) and \emph{Crowdsourcing} based approach to create and update curricula for individual learners. We show the design of this curriculum development system prototype, in which contributors receive AI-based recommendations to be able to define and update \emph{high-level learning goals}, \emph{skills}, and \emph{learning topics} together with associated learning content. This curriculum development system was also integrated into our personalized online learning platform. To evaluate our prototype we compared experts' opinion with our system's recommendations, and resulted in 89\%, 79\%, and 93\% F1-scores when recommending skills, learning topics, and educational materials respectively. Also, we interviewed eight senior level experts from educational institutions and career consulting organizations. Interviewees agreed that our curriculum development method has high potential to support authoring activities in dynamic, personalized learning environments.  
\end{abstract}



%\keywords{Learning Analytics, Lifelong Learning, Crowdsourcing, Artificial Intelligence}

\maketitle

\section{Introduction}
Rapid changes in the society such as high volatility in labor market demands \cite{vom2020labor}, the COVID situation around the world, and the growing need for personalization in learning environments \cite{shearer2020students,alamri2020using} have led to an increased attention to online informal learning \cite{lockee2021online,zhang2020understanding}. Nevertheless, educational systems that are capable to handle wide range of learners' context and requirements in these informal learning environments are still in their infancy \cite{berman2020critical}.

Although there have been many attempts to create personalized educational systems that meet learners' individual needs \cite{rojas2022personalized,tavakoli2020labour}, \emph{scalability} is still one of the most significant problems in this domain. This scalability problem leads to a situation that educational systems focus on specific key content domains (which can be manageable through a regular updating process) \cite{mousavinasab2021intelligent,tavakoli2020oer}, and/or become less sensitive about the quality of educational content (management of personal learning pathways, content quality, etc.) they offer \cite{zhang2020understanding}. % zheng2019personalized
We need a scalable dynamic curriculum development approach in order to be able to offer personalised education to learners. Such an approach should consider learners' needs (e.g. their needs towards labour market), capture relevant knowledge areas, include high-quality educational content for each area, and also be maintained with minimum efforts.

In this paper, we propose a curriculum development system to tackle the above mentioned challenges of online, personalized, informal learning environments using \emph{Artificial Intelligence (AI)} and \emph{Crowdsourcing} approaches. We defined and implemented four components with respective services including the crowd management service in our platform to empower curriculum developers when creating and maintaining a specific curriculum.

To validate our proposed system, we integrated it into our personal learning environment platform project, \emph{eDoer}\footnote{\url{www.edoer.eu}}, which aims to provide labour market driven personalized learning pathways and content. Subsequently, we calculated the accuracy of our proposed recommendation system by comparing it with pre-defined, data science related learning content (the domain we selected for evaluating our system) by three experts. Moreover, we interviewed eight education experts to provide us feedback on our objectives and the usability of our system. 

%In this paper, we will first show the most recent efforts on \emph{curriculum development} in Section \ref{sec-related}. Afterwards, in Section \ref{sec-method}, we explain our steps leading to the proposed curriculum development system prototype. Section \ref{sec-validation} will share the results of the initial evaluation of our proposed system and finally in Section \ref{sec-conclusion}, we will conclude our efforts and illustrate our next steps.

\section{State of the Art}\label{sec-related} 
Recent literature shows that there have been many attempts to overcome problems of curriculum development in online, informal education environments. We categorized these attempts based on the methodology they used in the following sections. At the end, we explain the lessons we learned to conclude our literature review.

\subsection{Artificial Intelligence Based Curriculum Development}
AI has been aiding curriculum development predominantly with using \emph{Machine Learning}\cite{somasundaram2020curriculum, tavakoli2020quality} and \emph{Text Mining}\cite{pattanshetti2018open, molavi2020extracting} methods. \cite{somasundaram2020curriculum} proposed an educational program model (i.e. prerequisite, content, expected outcome) based on labour market demand using \emph{AI back-propagation} concept in order to help learners to up-skill themselves towards their current or desired job. Although they claim that the result of their curriculum model is promising, they focused on the single area of \emph{Internet of Things} (\emph{IoT}). \cite{pattanshetti2018open} created knowledge graphs of \emph{Open Educational Resources} (\emph{OERs}) by applying \emph{Natural Language Processing} (\emph{NLP}) techniques to help authors to deliver their content to the proper audience. However, they only focused on content level and not on higher level learning goals. Moreover, some researches \cite{tavakoli2020quality, pattanshetti2021proposed} proposed approaches that perform automatic quality assessment for educational resources in order to help content providers to filter out low-quality content from their resources list. \cite{molavi2020extracting} introduces a novel method that uses \emph{Latent Dirichlet Allocation} (\emph{LDA}) \cite{jelodar2019latent} algorithm to extract topics covered by specific educational resources, in order to build learning pathways. Their application only focused on building curriculum for educating \emph{text mining} to learners.

\subsection{Crowdsourcing Based Curriculum Development}
As user participation in online learning platforms increases\cite{jiang2018review}, "the wisdom of the crowd" has a great potential for both teachers and learners. "Crowdsourcing for education" has been used for content-creation\cite{cross2014vidwiki} and also for sharing practical and theoretical knowledge\cite{pandey2017gut} in large scale. \cite{weld2012personalized} concludes that including crowd's opinion in the process of education is useful when it comes to building scalable and personalized curriculum. \cite{farasat2017crowdlearning} integrates crowdsourcing into pedagogical paradigm as "Crowdlearning". They suggest that including students into the creation of the curriculum not only increases the amount of content produced, but also improves the depth and performance of learning. However, crowdsourcing has its own flaws as well. Participation time and its effectiveness for complex tasks is one of the most noted issues\cite{stewart2010crowdsourcing, yang2021nonlinear}.

\subsection{Lessons we learned}
While crowdsourcing is generally useful for content-creation for curriculum development, and also contributes to a system's scalability, the time consumed by a participant can cause motivation problems and prevent their effective participation\cite{stewart2010crowdsourcing}. On the other hand, AI can help us to automate some tedious tasks such as quality assessment and learning topic extraction from educational resources. Therefore, we propose a novel curriculum development method, which utilizes the benefits of both AI and crowdsourcing approaches to generate dynamic and scalable personalized curricula.

The main research objectives of this paper are:
\begin{itemize}
    \item How can we empower educational service providers by using crowdsourcing to create, validate and maintain curricula?
    \item Does supporting curriculum designers and learning content authors with AI by recommending learning topics and high-quality educational resources for specific knowledge areas facilitate the process of developing up-to-date curriculum?
\end{itemize}

\section{Curriculum Development Framework}\label{sec-method}
We defined four main components for our framework: 1. \emph{high-level leaning goals} which consist of skills, 2. \emph{skills} which consist of learning topics, 3. \emph{learning topics} which consist of educational packages, and 4. \emph{educational packages} which include one or more educational resources. To help the contributors managing their content, each of these components is enhanced with the following services: 1. Add service for defining a component for the first time, 2. Suggestion service for collecting users' suggestions on a component and also automatically accept or reject them based on crowd opinions, 3. Recommendation service for providing insights for contributors based on existing open data (e.g. job vacancies, educational resources, and standard taxonomies) in order to help them adding and updating content.
Additionally, we designed a crowd management component to monitor and analyze contributors' activities and opinions when maintaining learning content.

At first we explain the details of designing and implementing our system, and subsequently show a first use case on an existing personalized educational platform.

\subsection{Managing High-level Learning Goals} \label{job}
\begin{figure}[h]
  \centering
  \includegraphics[width=.95\textwidth]{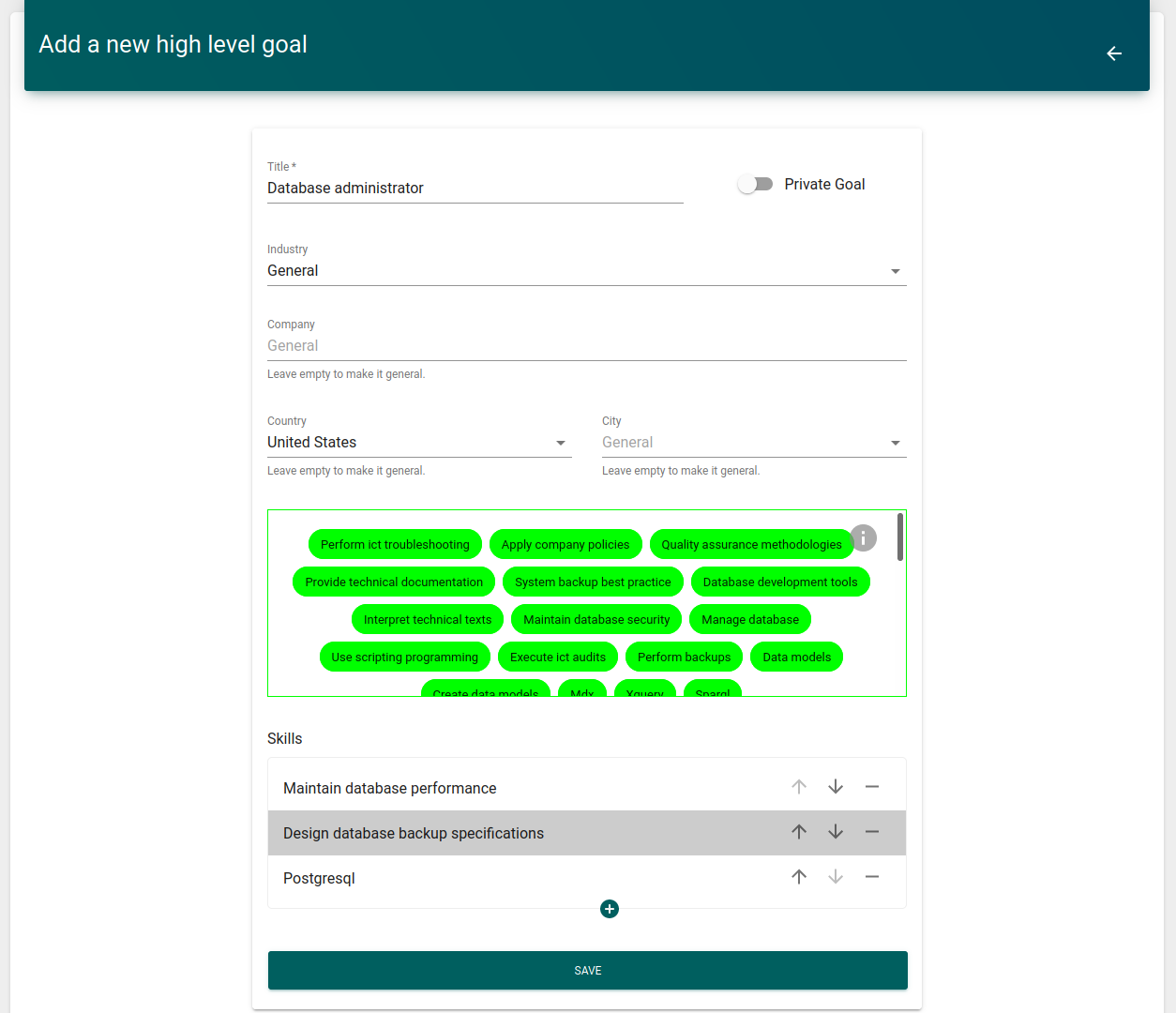}
  \vspace{-2pt}
  \caption{Screenshot of adding a high-level learning goal}
  \label{fig-add-goals}
\end{figure}

For adding high-level learning goals (consisting multiple skills), we collected \emph{titles} and optional \emph{descriptions} together with the following key contextual features to help learners \cite{tavakoli2020labour}: \emph{industry, company, city}, and \emph{country}. However, authors can decide not to contextualise learning objectives by keeping each of these fields as \emph{General}. Afterwards, our system recommends a list of goal related skills based on the title of the high-level goal, in order to capture necessary skills to master for the given learning goal. For these recommendations, we rely on the \emph{ESCO} \footnote {European Skills/Competences, qualifications, and Occupations: \url{https://ec.europa.eu/esco/portal/home}} data-set, which is a European multilingual classification system of skills, competences and occupations. ESCO is continuously updated by subject matter experts of the European Commission and includes 13,485 skills, linked to 2942 occupations. We match the title of the high-level goal with existing occupations in \emph{ESCO} using the \emph{Bleu score} \cite{papineni2002bleu} concept. We decided to use \emph{Bleu score} as it allows us to capture the closest term (occupation in this case) in the \emph{ESCO} data-set no matter in what language the title is. Finally, we recommend a list of skills linked to the closest occupation to the author. Authors can either select skills from the list, or add new skills manually. After finalizing the skill list, the author can sort skills based on the order they need to be shown to learners. Figure \ref{fig-add-goals} shows an screenshot about adding a high-level goal to our system.

After adding a high-level goal, users can view the page\footnote{Figure: \url{https://raw.githubusercontent.com/ali-faraji90/edoer/main/LAK22/Goal.png}} of the newly added high-level goal, which includes:
\begin{itemize}
    \item \emph{\textbf{The list of skills associated with the goal.}}
    \item \emph{\textbf{Opinion of other users (crowd) regarding the importance of a particular skill in relation to this goal.}} At each skill, users can see the importance of the particular skill for that particular high-level goal, based on up and down-votes (see \ref{voting}) of the crowd.
    \item \emph{\textbf{The suggestion list.}} After defining a high-level goal for the first time, editing can be done by addressing the crowd opinion. Therefore, all users can suggest: 1. adding a skill, 2. deleting a skill, 3. reordering skills for each high-level goal. To provide insights for users and to keep each goal updated, our system recommends other, potentially related skills using the aforementioned algorithm. After adding a suggestion, other users can provide their opinion about a particular suggestion by up and down-voting. Our system captures these actions and suggests a decision whether to reject or accept the suggestion (see \ref{crowd})
\end{itemize}

% \begin{figure}[h]
%   \centering
%   \includegraphics[width=\textwidth]{figures/Goal.png}
%   \vspace{-2pt}
%   \caption{Screenshot from a high-level learning goal page}
%   \label{fig-goal}
% \end{figure}

\subsection{Managing Skills} \label{skill}
\subsubsection{Defining Skills}
\begin{figure}[h]
  \centering
  \includegraphics[width=.95\textwidth]{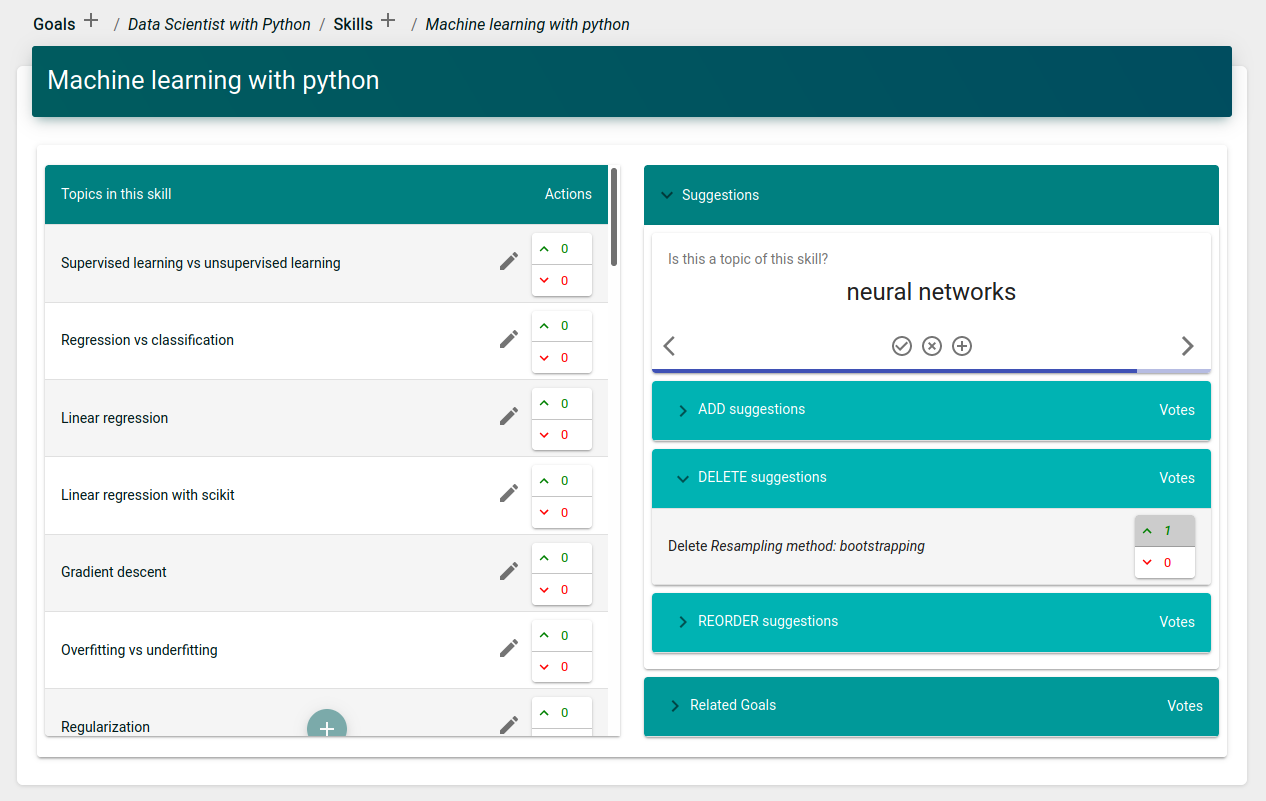}
  \vspace{-2pt}
  \caption{Screenshot from a skill page}
  \label{fig-skill}
\end{figure}

In order to define a skill, users receive string auto-completions as they are typing the title of a skill. This function is based on existing skills in the \emph{ESCO} standards. Auto-completion not only facilitates skill title definition, but also helps users to refrain from adding different titles for a single skill. After filling out the title and description fields, our system uses \emph{Youtube} playlists in order to provide insight for contributors regarding existing \emph{learning topics} on any new skill. The system automatically searches the skill title, collects related videos, and extracts the most important key-words by applying \emph{TF-IDF} \cite{ramos2003using} algorithm on the the collected video titles\footnote{Figure: \url{https://raw.githubusercontent.com/ali-faraji90/edoer/main/LAK22/Add-skill.png}}. The author can either select from this AI generated recommendation list or type in learning topics manually. Again, after finalizing the topic list, the author can sort the topics into an order of importance for learners. As depicted on Figure \ref{fig-skill} after defining the skill, a skill page will be created including: 1. the list of learning topics, 2. crowd opinion about the learning topic's importance (see \ref{voting}), 3. the suggestion list including \emph{system recommended learning topics} (see \ref{edit-skill}), \emph{adding, deleting,} and \emph{reordering} suggestions, which are monitoring our suggestion reviewing process (see \ref{review}), and 4. high-level learning goals associated with this skill.

% \begin{figure}[h]
%   \centering
%   \includegraphics[width=\textwidth]{figures/Add_skill.png}
%   \vspace{-2pt}
%   \caption{Screenshot of adding a skill}
%   \label{fig-add-skill}
% \end{figure}

\subsubsection{Editing Skills}\label{edit-skill}
After adding a skill, similarly to high-level goals, the crowd has the ability to edit its content. In order to contribute to the skill updating process, we provide some learning topic recommendations that other users can easily add as a suggestion (see Figure \ref{fig-skill}). These topic recommendations are generated as follows:
\begin{enumerate}
    \item \emph{\textbf{Collecting learning topics.}} We search each new learning topic title on \emph{Youtube}\footnote{using the \emph{Pafy} library: https://pypi.org/project/pafy/}, and collect at least 50 educational video transcripts labelled with that particular topic title. This exercise results in a two-column data-set including transcripts of educational resources in the first column, and 2. their associated learning topic in the second column.
    \item \emph{\textbf{Text pre-processing.}} On each collected transcript, we apply standard text pre-processing steps including \emph{converting to lower case, removing special characters, stemming,} and \emph{adding n-grams (bi/tri-grams)}. 
    \item \emph{\textbf{Applying labelled topic modeling.}} We use \emph{LLDA} (Labelled Latent Dirichlet Allocation) to extract topics from transcripts. \emph{LLDA} is a supervised version of \emph{LDA} \cite{jelodar2019latent}, which is a generative probabilistic topic model aiming at extracting the most important keywords for each topic labels. We collect the ten most important keywords, their probabilities for each topic and store them in the \emph{recommendation list}
    \item \emph{\textbf{Collecting content for skills.}} The \emph{recommendation list} we generated through topic modelling of learning topics is usually not exhaustive enough, as some skill related topics may not be represented well on the list. Therefore, we also perform search the title of the given skill on \emph{Youtube} playlists and collect at least 200 videos for each skill. 
    \item \emph{\textbf{Applying \emph{LDA} model on skill related learning resources.}} As for these resources we do not have any topic labels, we use an \emph{LDA} model, an unsupervised machine learning method, which considers each transcript as a distribution of different topics, each topic as a distribution of different words, and aims to extract existing topics together with their distribution of words. To find the number of learning topics related to a skill, which is the input of the \emph{LDA} model (as parameter $k$), we calculate $C_V$ \emph{Coherence} \cite{roder2015exploring} for different number of topics (between 2 to 50), and set $k$ with the topic amount that results in the highest coherence value. Again, for each newly extracted topic, we extract the ten most important words (and their probabilities) and add them to the previous words in our \emph{recommendation list}.
    \item \emph{\textbf{Removing the existing topics.}} In order to prevent recommending learning topics that are already associated with a skill, we remove those, already associated, topics from our recommendation list.
    \item \emph{\textbf{Sorting the recommendation list.}} To capture the most relevant topics quickly, we sort the words on the recommendation list according to their probability values in our topic models, and start recommending from the first item on the list.    
\end{enumerate}

\subsection{Managing Learning Topics} \label{topic}
Users can add learning topics by defining their title and description first. Afterwards, a topic page is created and contributors can start adding educational materials to a given topic. To facilitate adding an educational package to each topic, we collect a list of educational materials from \emph{Youtube} and \emph{Wikipedia}\footnote{Using Wikipedia library: \url{https://pypi.org/project/wikipedia/}} based on the title of the topic. Afterwards, each collected educational material passes through the following automatic quality control process:
\begin{enumerate}
    \item \emph{\textbf{Metadata based quality control.}} \cite{tavakoli2021metadata} showed a close relationship between the metadata quality and the content quality of educational resources. Accordingly, they built an openly available quality prediction model that detects the quality of educational materials based on their metadata. We adopted this approach to filter out low quality educational materials. Furthermore, based on this quality prediction model, we also calculated a quality score, which is shown to users for each recommended educational resource.
    \item \emph{\textbf{Content based quality control.}} To predict the relevancy of a particular educational resource to a learning topic, and, at the same time, to filter out irrelevant educational resources, we use the probability values of each word in the \emph{LLDA} model discussed in the previous section (see \ref{edit-skill}). 
\end{enumerate}

As recommended educational resources go through this automatic quality control process, users can directly associate those resources to learning topics. However, authors can still provide suggestions, for instance decoupling an educational resource from a topic, when they think it is not an appropriate (not relevant or it has low quality) content\footnote{Figure: \url{https://raw.githubusercontent.com/ali-faraji90/edoer/main/LAK22/Topic.png}}. Moreover, users can also have an overview of skills that require the knowledge of that particular learning topic.

% \begin{figure}[h]
%   \centering
%   \includegraphics[width=.95\textwidth]{figures/Topic.png}
%   \vspace{-2pt}
%   \caption{Screenshot from a learning topic page}
%   \label{fig-rates}
% \end{figure}

\subsection{Managing Educational Packages}
Users can also define educational packages including one or more educational resources. To do this, firstly, users need to name a title and an optional description for an educational package. Afterwards, they can add one or more educational resources either by importing from a \emph{URL}, or 2. uploading an educational resource. Subsequently, contributors can fine tune and set properties (i.e. title, description, format type, estimated time needed to complete the resource, source of the content, includes an example/theory or not, level of details, and if it is a recording of a class-based instruction or not) for each of the resources in the package. It should be mentioned that in order to facilitate the property setting process, we implemented a property extractor component which retrieves information from \emph{Youtube} educational videos and \emph{Wikipedia}. Finally, our system recommends related learning topics to contributors (using the LLDA model for the topics (see \ref{edit-skill})) in order to help them set the learning topics that are covered by the new educational package.

\subsection{Crowd's Opinion Management}\label{crowd}
In this part, we will explain our \emph{voting system, how contributors receive points}, and the \emph{process of suggestions' reviewing}.

\subsubsection{Voting System}\label{voting}
Users can up and down-vote skills and topics based on their perceived importance to their containing component. Therefore, with this feature, contributors are able to show how important 1. a skill for a high-level goal, 2. a learning topic for a skill, and 3. an educational package for a learning topic. Also, users can up/down-vote new suggestions (i.e. adding, deleting, and reordering suggestions) and give their achieved points in the context of the suggestion (see \ref{achieve_point}) to help a suggestion be accepted or rejected. By using this mechanism of voting, the system puts more weight on reliable users in the suggestion reviewing process (see \ref{review}). 

\subsubsection{Achieving Points} \label{achieve_point}
Contributors can collect points on each skill and also each learning topic. They can collect a point if their defined skills and topics receive an up-vote or if they are added as a learning goal to others' profile. Moreover, if the educational material they added receives up-votes from learners, they also receive a point for the target topic(s) of that educational material.

\subsubsection{Reviewing Process of Suggestions}\label{review}
Each and every suggestion needs to receive a minimum number of points to be approved in the system (currently ten points, but it can be changed in our configuration files) within a predefined period (currently a week, however it can also be changed in our configuration files), otherwise the suggestion will be automatically rejected. When a suggestion receives the minimum required points for approval, the system calculates the rate of positive received points to the all received points. If the rate is greater than 75\% (again customizable in our configuration), the suggestion is automatically accepted, otherwise rejected. 

\subsection{Use-case: Personalized, Goal-Driven Learning Recommendations}
\begin{figure}[h]
  \centering
  \includegraphics[width=.95\textwidth]{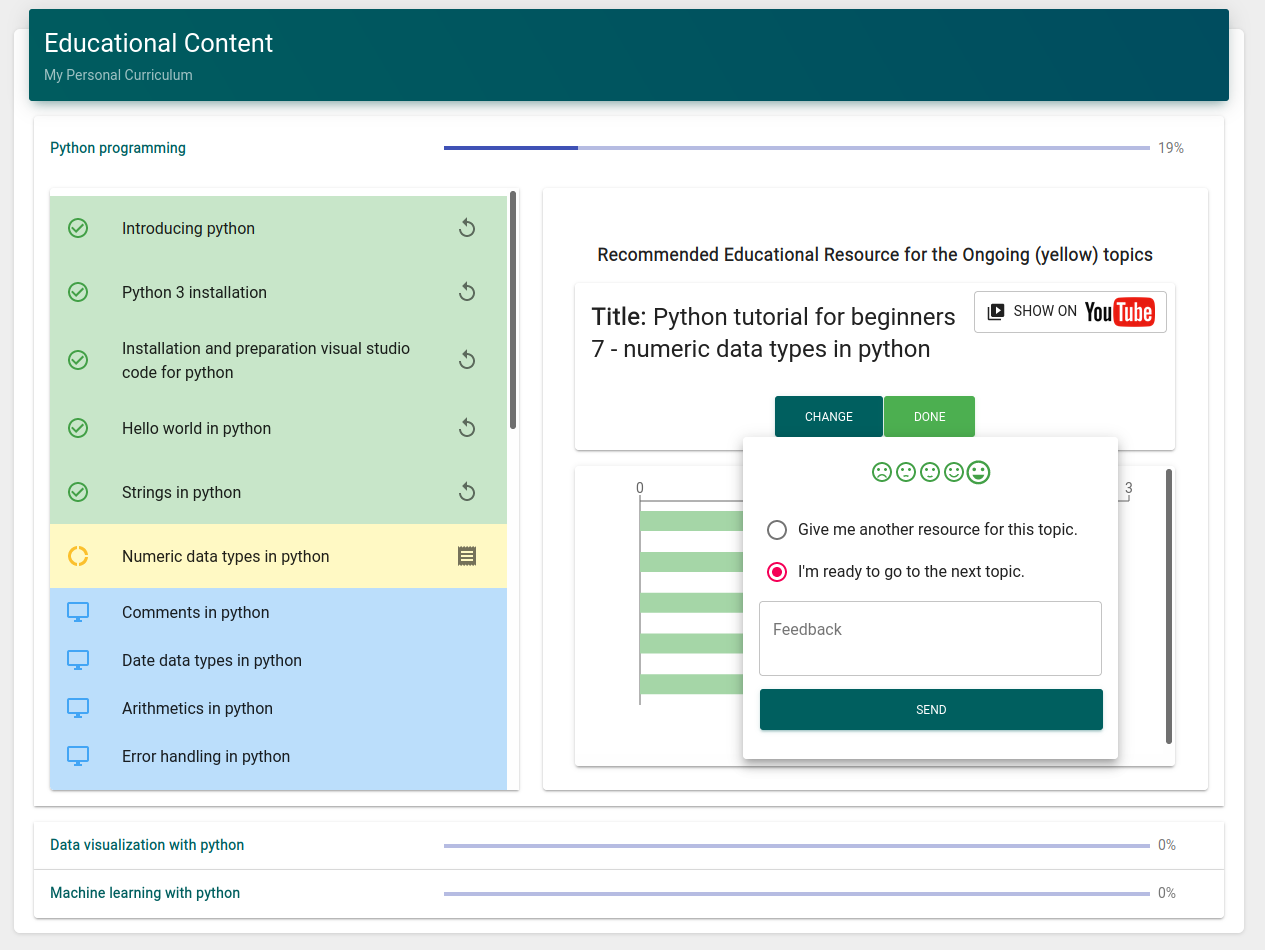}
  \vspace{-2pt}
  \caption{Screenshot from the personalized learning platform}
  \label{fig-curr}
\end{figure}

Our personalized learning system, \emph{eDoer}\footnote{\url{www.edoer.eu}}, is a platform in which learners 1. set their learning goals, 2. receive skill lists related to their target goals, 3. select skills from the list, what they want to master, 4. generate their learning dashboard, which includes all selected skills and associated learning topics, and 5. receive personalized educational resources based on their preferences and behavior history on the platform. Figure \ref{fig-curr} shows learners’ dashboard in our personalized learning content recommendation system. We used this educational platform as the first use case for our AI-based crowdsourcing system, to investigate on one hand how this technology aids contributors and authors to define and maintain educational content, and on the other hand, how does it empower learners by providing up-to-date personalized curricula. 

\section{Validation}\label{sec-validation}
To evaluate the system, we decided to measure the accuracy of our recommendations during the authoring process in the domain of \emph{Data Science}. Moreover, we designed a semi-structured subject matter expert interview protocol, to evaluate the authoring process with users. In this section, we will showcase the results of these efforts.

\subsection{Recommendation Accuracy}
To evaluate our recommendations, we asked three experts with at least five years of academic and ten years of industrial experience in the area of \emph{Data Science} to manually (without using our platform) specify the following high-level goals: \emph{Data Scientist}, \emph{Data Analyst}, and \emph{Business Analyst} together with related skills and learning topics. Subsequently, they used our system to define these jobs with our skill recommendations (see \ref{job}). 
The list of skills produced by the experts were then compared with the list of recommendations of our system. The results showed that our system, on average, had an F1-score of 89\% when it comes to recommending relevant skills for high level goals. When it comes to learning topic recommendation for data science related skills (see \ref{skill}) our evaluation resulted in the following F1-scores: \emph{Python programming}: 83\%, \emph{Machine learning}: 76\%, \emph{Statistics}: 79\%, \emph{Data visualization}: 75\%, which meant 79\% as weighted average. 

Finally, when recommending specific learning content for learning topics (see \ref{topic}), experts examined their validity. Only if the recommended content was marked as high-quality and relevant to the topic, we considered it as a valid recommendation. This evaluation revealed that our educational content recommendation method provided high-quality relevant materials in 93\% of time\footnote{As different curricula are continuously developed in our system, to receive an up-to-date data-set of the existing components and recommendations, contact the authors.}.

\subsection{Subject Matter Evaluation}
In order to evaluate the proposed authoring process, we designed and executed interviews with 8 senior members (i.e. managers, professors, associate professors, and researchers) from different organizations: Participant\_1 from \emph{Ericsson} company in Sweden, Participant\_2 \emph{University of Amsterdam}, Participant\_3 \emph{KU Leuven University}, Participant\_4 from \emph{TIB - Leibniz Information Centre for Science an Technology}, Participant\_5 from \emph{University of Bonn}, Participant\_6 from {Netherlands AI Coalition}, Participant\_7 from \emph{the American Psychological Association}, Participant\_8 from \emph{Career and Life Planning (CALP)}. The structure of the interviews was as follows: 1. Introducing the system and its logic (\textasciitilde 20 minutes), 2. using the authoring system and its features for about 20 minutes, and 3. going through a semi-structured interview with the assistance of a questionnaire\footnote{The questionnaire is available here: \url{https://forms.gle/A17KHhWGoUH9WFuc8}} (\textasciitilde 20 minutes).

Participant\_1 and participant\_2 mentioned that providing an environment in which learners can be informed about when and how to build their careers is the most important part of the system. Also, Participant\_3 considered providing insights for authors regarding the relevant skills, learning topics, and educational materials as a key feature of our system. Participant\_4, Participants\_5 and Participant\_7 emphasized that combining AI with crowdsourcing can improve the quality of each recommendation component, and therefore this is the most promising function of the system. However, they suggested to provide as much clarity as possible for users on how our \emph{AI} and \emph{crowd-management} components collaborate with each other. Participant\_6, besides pointing at the usefulness of our system when it comes to matching jobs and their required knowledge dynamically, mentioned that integrating the ability for testing a knowledge should be part of our future steps. Participant\_8 believed goals page and curriculum are the most important elements of our system and thought enriching these parts should have a priority in our future work.

Ultimately, \emph{100\%} of the interviewees agreed that "creating dynamic personalized curricula for learners" as our objective is extremely important and timely. Also, only \emph{12.5\%} (1 out of 8) was unsatisfied with the usability of our prototype system, which shows that most of the participants consider our prototype usable already. 

\section{Conclusion and Future Work}\label{sec-conclusion}
Throughout this paper, we showcased a novel learning content authoring system which helps authors to build personalized curricula for learners. By providing intelligent recommendations, this system aids contributors to define 1. high-level learning goals consisting of skills, 2. skills built by learning topics, and 3. learning topics with related educational materials. We believe that such a system not only helps authors to define and maintain their educational content, but also empowers learners through setting their own learning objectives, receiving personalized recommendations, and being up-to-date on desirable knowledge. Evaluating our recommendations in the context of the data science showed that our system can provide 89\% F1-accuracy in matching high-level goals and their skills, 79\% F1-accuracy matching skills and their learning topics, and 93\% precision in recommending high-quality, relevant educational materials. Moreover, we validated the main objective and usability of our system by interviewing eight subject matter experts in the area of education which showed that they were satisfied with the objective and usability of the our proposed system.

\bibliographystyle{ACM-Reference-Format}
\bibliography{paper}

\end{document}